\documentclass[3p, twocolumn]{elsarticle}

\usepackage[breaklinks=true]{hyperref}
\usepackage[anythingbreaks]{breakurl}
\usepackage{amsthm}
\usepackage{amsmath}
\usepackage{amsfonts}
\usepackage{amssymb}
\usepackage{amsfonts}
\usepackage{epsfig}
\usepackage{bbm}
\usepackage{todonotes}
\usepackage{soul}
\usepackage{nicefrac}
\usepackage[capitalize]{cleveref}
\usepackage{booktabs}
\usepackage[super]{nth}
\usepackage{float}
\usepackage{caption}
\usepackage{subcaption}
\usepackage{dblfloatfix}
\usepackage{xcolor}

\usetikzlibrary{arrows.meta, 
	bending, 
	calc, chains, 
	decorations.pathmorphing,
	positioning
}

\usepackage{pgfplots}

\usepgfplotslibrary{fillbetween}
\usetikzlibrary{intersections}

\newcommand*\td{\mathop{}\!\mathrm{d}}

\definecolor{rwth}   {RGB}{  0  84 159}
\definecolor{rwth-75}{RGB}{ 64 127 183}
\definecolor{rwth-50}{RGB}{142 186 229}
\definecolor{rwth-25}{RGB}{199 221 242}
\definecolor{rwth-10}{RGB}{232 241 250}

\definecolor{black}   {RGB}{  0   0   0}
\definecolor{black-75}{RGB}{100 101 103}
\definecolor{black-50}{RGB}{156 158 159}
\definecolor{black-25}{RGB}{207 209 210}
\definecolor{black-10}{RGB}{236 237 237}

\definecolor{magenta}   {RGB}{227   0 102}
\definecolor{magenta-75}{RGB}{233  96 136}
\definecolor{magenta-50}{RGB}{241 158 177}
\definecolor{magenta-25}{RGB}{249 210 218}
\definecolor{magenta-10}{RGB}{253 238 240}

\definecolor{yellow}   {RGB}{255 237   0}
\definecolor{yellow-75}{RGB}{255 240  85}
\definecolor{yellow-50}{RGB}{255 245 155}
\definecolor{yellow-25}{RGB}{255 250 209}
\definecolor{yellow-10}{RGB}{255 253 238}

\definecolor{petrol}   {RGB}{  0  97 101}
\definecolor{petrol-75}{RGB}{ 45 127 131}
\definecolor{petrol-50}{RGB}{125 164 167}
\definecolor{petrol-25}{RGB}{191 208 209}
\definecolor{petrol-10}{RGB}{230 236 236}

\definecolor{turkis}   {RGB}{  0 152 161}
\definecolor{turkis-75}{RGB}{  0 177 183}
\definecolor{turkis-50}{RGB}{137 204 207}
\definecolor{turkis-25}{RGB}{202 231 231}
\definecolor{turkis-10}{RGB}{235 246 246}

\definecolor{grun}   {RGB}{ 87 171  39}
\definecolor{grun-75}{RGB}{141 192  96}
\definecolor{grun-50}{RGB}{184 214 152}
\definecolor{grun-25}{RGB}{221 235 206}
\definecolor{grun-10}{RGB}{242 247 236}

\definecolor{maigrun}   {RGB}{189 205   0}
\definecolor{maigrun-75}{RGB}{208 217  92}
\definecolor{maigrun-50}{RGB}{224 230 154}
\definecolor{maigrun-25}{RGB}{240 243 208}
\definecolor{maigrun-10}{RGB}{249 250 237}

\definecolor{orange}   {RGB}{246 168   0}
\definecolor{orange-75}{RGB}{250 190  80}
\definecolor{orange-50}{RGB}{253 212 143}
\definecolor{orange-25}{RGB}{254 234 201}
\definecolor{orange-10}{RGB}{255 247 234}

\definecolor{rot}   {RGB}{204   7  30}
\definecolor{rot-75}{RGB}{216  92  65}
\definecolor{rot-50}{RGB}{230 150 121}
\definecolor{rot-25}{RGB}{243 205 187}
\definecolor{rot-10}{RGB}{250 235 227}

\definecolor{bordeaux}   {RGB}{161  16  53}
\definecolor{bordeaux-75}{RGB}{182  82  86}
\definecolor{bordeaux-50}{RGB}{205 139 135}
\definecolor{bordeaux-25}{RGB}{229 197 192}
\definecolor{bordeaux-10}{RGB}{245 232 229}

\definecolor{violett}   {RGB}{ 97  33  88}
\definecolor{violett-75}{RGB}{131  78 117}
\definecolor{violett-50}{RGB}{168 133 158}
\definecolor{violett-25}{RGB}{210 192 205}
\definecolor{violett-10}{RGB}{237 229 234}

\definecolor{lila}   {RGB}{122 111 172}
\definecolor{lila-75}{RGB}{155 145 193}
\definecolor{lila-50}{RGB}{188 181 215}
\definecolor{lila-25}{RGB}{222 218 235}
\definecolor{lila-10}{RGB}{242 240 247}

\begin{document}

\begin{frontmatter}
	\title{Self-Adaptive Physics-Informed Quantum Machine Learning for Solving Differential Equations}
    
    \author[a,b,c]{Abhishek Setty\corref{mycorrespondingauthor}}
    \cortext[mycorrespondingauthor]{Corresponding author}
    \ead{a.setty@fz-juelich.de}  
    \author[a]{Rasul Abdusalamov}
    \author[b,c]{Felix Motzoi}
    
    \address[a]{Department of Continuum Mechanics, RWTH Aachen University, Germany}
    \address[b]{Forschungszentrum Jülich, Institute of Quantum Control (PGI-8), D-52425 Jülich, Germany}
    \address[c]{Institute for Theoretical Physics, University of Cologne, D-50937 Cologne, Germany}
    
\begin{abstract}	
 Chebyshev polynomials have shown significant promise as an efficient tool for both classical and quantum neural networks to solve linear and nonlinear differential equations. In this work, we adapt and generalize this framework in a quantum machine learning setting for a variety of problems, including the 2D Poisson's equation, second-order linear differential equation, system of differential equations, nonlinear Duffing and Riccati equation. In particular, we propose in the quantum setting a modified Self-Adaptive Physics-Informed Neural Network (SAPINN) approach, where self-adaptive weights are applied to problems with multi-objective loss functions. We further explore capturing correlations in our loss function using a quantum-correlated measurement, resulting in improved accuracy for initial value problems. We analyse also the use of entangling layers and their impact on the solution accuracy for second-order differential equations. The results indicate a promising approach to the near-term evaluation of differential equations on quantum devices.
	\end{abstract}

\end{frontmatter}
\thispagestyle{empty}

\section{Introduction}\label{sec1}
With the invention of calculus, a new era of mathematics was created by introducing differential equations (DEs) by Sir Isaac Newton and Gottfried Leibniz \cite{newton1774methodus, leibniz1858historia}. DEs are ubiquitous in describing physical phenomena, beginning with classical mechanics, fluid dynamics, electromagnetism, and quantum mechanics. To date, the search for efficient and accurate solutions to DEs has remained a challenging task and many problems have no closed-form solution. Numerical methods have been introduced in the last century, such as the finite element method, finite differences, or finite volume method that can be used to solve more complex systems \cite{courant1928partiellen, galerkin1915rods, ritz1909neue, demirdzic1988numerical}. These methods remain crucial for many engineering applications and are further investigated. Nevertheless, they also have limitations, such as the problem with discretization errors, convergence rates, and the requirement for large computational resources.

With the recent rise in machine learning (ML) algorithms, a novel concept was introduced by Raissi et al., where physical principles were combined with data-driven techniques to solve partial differential equations \cite{raissi2017physics}. The key idea is to implement the DE together with the physical constraints of the system as additional loss terms during the neural network training. The introduced neural networks are known as physics-informed neural networks (PINNs) and have been used for a variety of applications such as for solving inverse problems involving nonlinear partial differential equations \cite{RAISSI2019686}, problems in fluid mechanics \cite{cai2021physics} or for the analysis of defects in materials \cite{Zhang2022}. PINNs can handle noisy or incomplete data and do not rely on any fine-grained numerical discretization. Although PINNs have a promising future, one major problem remains: training the networks is related to immense computation costs. 

Meanwhile, recent breakthroughs in quantum computing have suggested computational advantages in machine learning and solving linear systems of equations. The seminal Harrow-Hassidim-Lloyd algorithm (HHL) \cite{lloyd2014quantum, lloyd2010quantum} provides a system solution that is encoded into the amplitude of a quantum state through quantum phase estimation and amplitude amplification. This method has been extended to solve nonlinear DEs in computational fluid dynamics and structural mechanics \cite{lapworth2022hybrid, tosti2022review}. An alternative method for solving DEs also {using amplitude encoding} employs instead employing variational quantum circuits \cite{lubasch2020variational, umer2024nonlinear}, where a quantum nonlinear programming unit (QNPU) may efficiently calculate nonlinear product operators of the field, even when these nonlinearities are large. A drawback of this method is however that error can propagate iteratively in the current step depending on the previous step solution while having to deal also with relatively deep (though narrow) circuits. 
Lastly, quantum kernel methods have been proposed to solve regression as well as differential equations \cite{paine2023quantum}, leading to an advantageous choice for the loss function that allows for efficient multi-point training. Additional advances have been made in resolving partial differential equations (PDEs) in the context of quantum amplitude estimation, employing Chebyshev points \cite{oz2023efficient}. The resolution of advection-diffusion equations through the use of variational quantum linear systems has been introduced in \cite{demirdjian2022variational}. 

Recently, a different approach has been proposed that avoids problems from amplitude encoding by using feature map encoding to approximate functions and solve nonlinear first-order differential equations using quantum circuits \cite{mitarai2018quantum, kyriienko2021solving, williams2023quantum}. The approach encodes the state using a latent space representation defined in the high-dimensional Hilbert space, benefiting from the large expressive power of quantum feature maps and parameterized quantum circuits. 

In our work, we extend the usage of the quantum Chebyshev feature map while solving a variety of differential equations. In addition, we incorporate the approach of self-adaptive weights for balancing multi-objective loss functions while solving differential equations. Particularly, we modify the method of self-adaptive weights introduced for classical PINNs \cite{mcclenny2023self} into the setting of the proposed variational quantum algorithm. Moreover, we study the use of correlated measurement observables and more entangling layers for better convergence. Our results show that these approaches lead to a measurable improvement in the convergence and accuracy for solving the given test cases, namely the Duffing equation, Riccati equation, system of differential equations, second-order linear differential equation, and the 2D Poisson's equation. 

\section{Methods}\label{sec:methods}
\begin{figure}[t]
\centering
\includegraphics[width=\columnwidth]{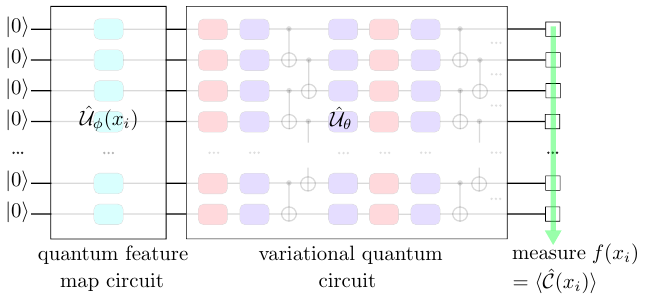}
\captionsetup{width=\columnwidth}
\caption[Quantum circuit with a feature map, variational block, and measurement.]{Quantum circuit for solving differential equations. On the left, a feature map ($\phi(x_i)$) block is used for encoding the value of a function at a specific input $x_i$, followed by a variational parameter block $\hat{\mathcal{U}}_{\theta}$, in which $\theta$ is a set of variational parameters optimized over cost function. In the end, the desired function $(f(x_i))$ in the differential equation is evaluated as an expectation value of the circuit based on a chosen operator $\hat{\mathcal{C}}$.} 
\label{fig:QC}
\end{figure}

The variational quantum circuit to solve differential equations proposed by Kyriienko et al. \cite{kyriienko2021solving} is presented in \cref{fig:QC}. In this circuit, there are three main parts, namely, a quantum feature map circuit for encoding Chebyshev polynomials, a variational circuit optimizing the cost function, and measurement (where we test correlated observables in the quantum loss functions). An overview of the variables used in this work is mentioned in \cref{tab:Variables} for the sake of clarity.

\begin{table}[ht]
	\centering
	\caption{Overview of used variables.}
	\label{tab:Variables}
	\begin{tabular}{p{1.cm}p{5cm}}
		\toprule
		\textbf{Variables} & \\
		\midrule
		$N_{f}$ & Set of collocation points \\
		$N_{b}$ & Set of boundary points \\
		$x_{f}^{i}$ & Collocation grid points \\
		$x_{b}^{i}$ & Boundary points \\
		$u$/$u_{\text{true}}$ & True solution \\
		$u_{\text{pred}}$ & Predicted solution \\
		$\alpha_{f}$ & Residual loss weight \\
		$\alpha_{b}$ & Boundary loss weight \\
		$\hat{\mathcal{U}}_{\phi}$ & Quantum feature map \\
		$\hat{\mathcal{U}}_{\theta}$ & Variational quantum circuit \\
		\bottomrule
	\end{tabular}
\end{table}

For a single independent variable $x_i$, the quantum state for this circuit will be $|f_{\phi, \theta} (x_i)\rangle = \hat{\mathcal{U}}_{\theta}\hat{\mathcal{U}}_{\phi}(x_i) |0 \rangle$. The real-valued classical function $f(x_i)$ is calculated by expectation value of a predefined Hermitian cost operator $\hat{\mathcal{C}}$, such that, $f(x_i) = \langle f_{\phi , \theta}(x_i) | \hat{\mathcal{C}} | f_{\phi, \theta}(x_i)\rangle$. Solving differential equations requires computing gradients of desired functions with respect to the independent variable, $\frac{\td f(x)}{\td x}$. For the given circuit, this derivative requires considering quantum feature map $(\phi(x))$, such as, $\frac{\td \hat{\mathcal{U}}_{\phi}(x)}{\td x} = \sum_j \hat{\mathcal{U}}_{\td \phi ,j}(x).$ The Chebyshev feature map which encodes Chebyshev polynomials into the circuit is given by 
\begin{equation}\label{eq:Chebyshev}
    \hat{\mathcal{U}}_{\phi}(x) = \bigotimes_{j=1}^N R_{Y,j}(2j \arccos{x}),    
\end{equation}
where $j$ is the qubit index. In-depth details of this approach are available in  \ref{sec:QFM}.

\subsection{Derivation of Second Order Derivative}\label{sec:second_derivative_section}
For solving second-order differential equations, we need to compute second-order derivatives of circuits. For a problem solved in \cref{subsec:SecondDE}, the second derivative $\frac{\td^2 u_{\text{pred}}}{\td(x_f^{i})^2}$ in \cref{eq:residual_second_order} considering the quantum Chebyshev feature map (refer \cref{eq:Chebyshev}) is calculated using chain rule as follows,
\begin{equation}\label{eq:d2u_dx2_derivation}
	\begin{aligned}
		\frac{\td^2u_{\text{pred}}}{\td (x_f^{i})^2} &= \frac{\td}{\td x_f^{i}} \left( \frac{\td u_{\text{pred}}}{\td x_f^{i}} \right), \\
		&= \frac{\td}{\td x_f^{i}} \left( \sum_j \frac{\td u_{\text{pred}}}{\td  \phi_j (x_f^{i})} \frac{\td \phi_j (x_f^{i})}{\td x_f^{i}} \right),\\
		&= \sum_{j, k} \frac{\td^2 u_{\text{pred}}}{\td \phi_j(x_f^{i}) \td \phi_k(x_f^{i})} \left[\frac{4jk}{1-(x_f^{i})^2}\right] \\
        &\quad + \sum_j \frac{\td u_{\text{pred}}}{\td \phi_j (x_f^{i})} \left[\frac{-2j x_f^{i}}{(1-(x_f^{i})^2)^{\frac{3}{2}}}\right].
	\end{aligned}
\end{equation}	
Notice that \cref{eq:d2u_dx2_derivation} contains a term where the denominator can become zero in case of $x_f^{i} = 1.0$ leading to numerical instability. Therefore in the problem of solving second-order DE (refer \cref{subsec:SecondDE}), the last collocation point is chosen with $x_{f}^{30} = 0.99$. Correspondingly, the Dirichlet boundary conditions are evaluated at this point.

\subsection{Self-Adaptive Weights for Multi-Objective Loss Function}
\label{section:SAPINNs}
In the weighted multi-objective loss function, the weights balancing the total loss function can be treated as variables instead of constants as shown in \ref{sec:weighted_objective}. The Self-Adaptive Physics-Informed Neural Networks \cite{mcclenny2023self} also referred to as SAPINNs, allow these weights to be trainable and they are applied to each training point individually. The idea behind SAPINNs is to maximize the weights and minimize the losses. We explore the applicability of this classical method for the variational quantum algorithm proposed in this work. If the variational parameters of the quantum circuit are given by $\boldsymbol{\theta}$, then the total loss function can be expressed by
\begin{equation}
	\mathcal{L}(\boldsymbol{\theta}, \boldsymbol{\lambda}_f, \boldsymbol{\lambda}_b) = \mathcal{L}_f(\boldsymbol{\theta}, \boldsymbol{\lambda}_f) +  \mathcal{L}_b(\boldsymbol{\theta}, \boldsymbol{\lambda}_b) \, ,
\end{equation}
where $\boldsymbol{\lambda}_f = (\lambda_f^1, \cdots, \lambda_f^{N_f})$ and $\boldsymbol{\lambda}_b = (\lambda_b^1, \cdots, \lambda_b^{N_b})$ are trainable, non-negative self-adaptation weights for the residual, and boundary points, respectively. These loss terms are given by
\begin{align}
	\mathcal{L}_f(\boldsymbol{\theta}, \boldsymbol{\lambda}_f) &= \frac{1}{N_f} \sum_{i=1}^{N_f} m(\lambda_f^{i})\left(f(x_f^{i}; \boldsymbol{\theta})\right)^2, \\
	\mathcal{L}_b(\boldsymbol{\theta}, \boldsymbol{\lambda}_b) &= \frac{1}{N_b} \sum_{i=1}^{N_b} m(\lambda_b^{i})\left(u_{\text{pred}}(x_b^{i}; \boldsymbol{\theta}) - u_b(x_b^{i})\right)^2,
\end{align}
where the self-adaptation mask function $m(\lambda)$ defined on $[0, \infty)$ is a non-negative, differentiable, strictly increasing function of $\lambda$. An interesting point to note about SAPINNs is that the loss $\mathcal{L}(\boldsymbol{\theta}, \boldsymbol{\lambda}_f, \boldsymbol{\lambda}_b)$ is minimized with respect to the variational parameters $\boldsymbol{\theta}$, as usual, but is maximized with respect to the self-adaptation weights $\boldsymbol{\lambda}_f, \boldsymbol{\lambda}_b$.  All parameters are updated by a gradient descent approach with 
\begin{align}
	\boldsymbol{\theta}^{k+1} &= \boldsymbol{\theta}^{k} - \eta^k \nabla_{\boldsymbol{\theta}} \mathcal{L}(\boldsymbol{\theta}^k, \boldsymbol{\lambda}_f^k, \boldsymbol{\lambda}_b^k) \, , \\
	\boldsymbol{\lambda}_f^{k+1} &= \boldsymbol{\lambda}_f^{k} + \rho_f^k \nabla_{\boldsymbol{\lambda}_f} \mathcal{L}(\boldsymbol{\theta}^k, \boldsymbol{\lambda}_f^k, \boldsymbol{\lambda}_b^k) \, , \\
	\boldsymbol{\lambda}_b^{k+1} &= \boldsymbol{\lambda}_b^{k} + \rho_b^k \nabla_{\boldsymbol{\lambda}_b} \mathcal{L}(\boldsymbol{\theta}^k, \boldsymbol{\lambda}_f^k, \boldsymbol{\lambda}_b^k) \, ,
\end{align}
where $\eta^k > 0$ is the learning rate for the variational parameters at step $k, \rho_p^k = 0.01$ is a separate learning rate chosen for the self-adaptation weights for $p = f, b$. Furthermore, the gradients are described as
\begin{equation}
\begin{split}
	\nabla_{\boldsymbol{\lambda}_f} \mathcal{L}(\boldsymbol{\theta}^k, &\boldsymbol{\lambda}_f^k, \boldsymbol{\lambda}_b^k) = \\ 
        &\frac{1}{N_f} \begin{bmatrix}
		m^{\prime}(\lambda_f^{k, 1})\left(f(x_f^{1}; \boldsymbol{\theta})\right)^2 \\
		\cdots\\
		m^{\prime}(\lambda_f^{k, N_f}) \left(f(x_f^{N_f}; \boldsymbol{\theta})\right)^2 \\
	\end{bmatrix} \, ,
 \end{split}
 \end{equation}
 \begin{equation}
\begin{split}
	&\nabla_{\boldsymbol{\lambda}_b} \mathcal{L}(\boldsymbol{w}^k, \boldsymbol{\lambda}_f^k, \boldsymbol{\lambda}_b^k) = \\ 
 &\frac{1}{N_b} \begin{bmatrix}
		m^{\prime}(\lambda_b^{k, 1}) \left(u_{\text{pred}}(x_b^{1}; \boldsymbol{\theta}) - u_b(x_b^{1})\right)^2 \\
		\cdots\\
		m^{\prime}(\lambda_b^{k, N_b}) \left(u_{\text{pred}}(x_b^{N_b}; \boldsymbol{\theta}) - u_b(x_b^{N_b})\right)^2 
	\end{bmatrix}.
\end{split}
\end{equation}

\subsection{Mask Functions for Self-Adaptive Weights} \label{sec:mask_functions}
The self-adaptive weights method works on employing mask functions for each loss term in the multi-objective loss function. These mask functions act as adaptive weights that penalize each loss term that is not converging during optimization. A mask function consists of variational parameters $\lambda$ which are optimized along with variational quantum circuit parameters. 

For the problem solved in \cref{subsec:SecondDE}, two masked functions $m$ have been tested: a polynomial mask (referred to as \textit{Polynomial}) and a logistic mask (referred to as \textit{Logistic}). Since the polynomial mask is subjected to numerical overflow, these masks are modified with threshold values for $\lambda$, over which the curve flattens (see \cref{fig:Problem3p2a}). The hyperparameters of polynomial mask $m(\lambda_{i}) = c\lambda_{i}^q$ for the residual and boundary conditions are heuristically chosen. All $\lambda$ are initialized to 1 and the threshold values $\lambda_{\text{max}}$ are chosen accordingly. For the polynomial mask functions, the values for residual are chosen as $c_r = 0.1$, $q_r = 2$, and $\lambda_{\text{max},r} = 10$. For the boundary conditions these parameters are provided with $c_b = 1000$, $q_b = 2$, and $\lambda_{\text{max},b} = 1000$. In the case of the logistic mask functions, the saturation value for the residual is chosen as $0.1$, and for boundary conditions, it is provided with $1000$.

\subsection{Solving Partial Differential Equation}\label{sec:methods_Poisson}
\begin{figure}[t]
\centering
\includegraphics[width=\columnwidth]{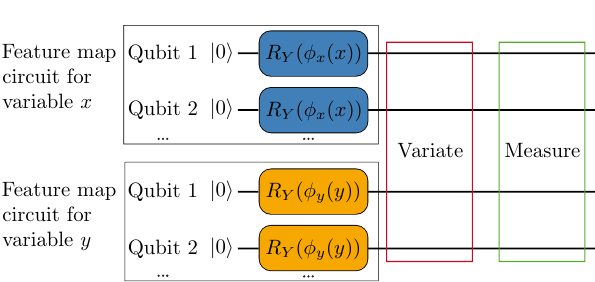}
\captionsetup{width=\columnwidth}
\caption{Visualization of a quantum circuit for encoding two independent variables using separate feature maps $\hat{\mathcal{U}}_{\phi_x}(x) = \bigotimes_{j=1}^{N_x} R_{Y,j}(\phi_x(x))$ and $\hat{\mathcal{U}}_{\phi_y}(y) = \bigotimes_{m=1}^{N_y} R_{Y,m}(\phi_y(y))$. The qubit numbering for each feature map starts from $1$ as they must be encoded as independent Chebyshev polynomials into the circuit. Also, the variational part of the quantum circuit (in red box) and measurement of the circuit (in green box) are represented following the feature map blocks.} 
\label{fig:Multivariable_circuit}
\end{figure}

Solving a partial differential equation requires encoding multiple independent variables into the quantum feature map. Therefore, a novel quantum circuit is needed, as shown in \cref{fig:Multivariable_circuit}. That is, the desired function $u(x, y)$ requires quantum feature mapping of two independent variables $x$ and $y$. Therefore, we define two unitary operators $\hat{\mathcal{U}}_{\phi_x}(x)$ and $\hat{\mathcal{U}}_{\phi_y}(y)$ for each nonlinear feature map $\phi_x(x)$ and $\phi_y(y)$. The combined unitary operator is written as,
	$\hat{\mathcal{U}}_{\phi_{x,y}}(x, y) = \hat{\mathcal{U}}_{\phi_x}(x) \otimes \hat{\mathcal{U}}_{\phi_y}(y)$.
The state of the quantum circuit for these two variables is given by
\begin{equation}
	|f_{\phi_{x, y}, \theta}(x, y)\rangle = \hat{\mathcal{U}}_{\theta}\hat{\mathcal{U}}_{\phi_{x, y}}(x, y)|0\rangle \ .
\end{equation} 
Therefore, the expectation value of the circuit with the Hamiltonian $\hat{\mathcal{C}}$ is given as,
\begin{equation}
	f(x, y) = \langle f_{\phi_{x, y}, \theta}(x, y)|\hat{\mathcal{C}}|f_{\phi_{x, y}, \theta}(x, y)\rangle.
	\label{eq:Problem4Output}		
\end{equation}
The feature maps in this study, $\phi_x(x)$ and $\phi_y(y)$ are both Chebyshev feature maps, $\phi_{xj}(x) = 2j \arccos{(x)}$ and $\phi_{ym}(y) = 2m \arccos{(y)}$, respectively. The unitary operator encoding the feature map for variable $x$ is described as $\hat{\mathcal{U}}_{\phi_x}(x) = \bigotimes_{j=1}^{N_x} R_{Y,j}(2j \arccos{x})$ (i.e.~as the top box in \cref{fig:Multivariable_circuit}). Similarly, for variable $y$ is described as $\hat{\mathcal{U}}_{\phi_y}(x) = \bigotimes_{m=1}^{N_y} R_{Y,m}(2m \arccos{x})$ (i.e.~as the bottom box in \cref{fig:Multivariable_circuit}). Here $N_x$ and $N_y$ are the number of qubits for each independent variable. The combined observable is the summation of Pauli-Z operators of all qubits such as $\hat{\mathcal{C}} = \sum_j Z_j + \sum_m Z_m$. 

For the problem solved in \cref{sec:poisson_results}, we considered a 2D Poisson's equation \cref{eq:Poisson2D}. The Laplace term in Poisson's equation \cref{eq:Poisson2D} considering the quantum Chebyshev feature maps (see \cref{eq:Chebyshev}) is given by,
\begin{equation}
	\begin{aligned}
		\Delta u(x, y) &=\sum_{j,k} \frac{\partial^2 u_{\text{pred}}}{\partial \phi_{xj}(x_f^{i})\partial \phi_{xk}(x_f^{i})}\left[\frac{4jk}{1-(x_f^{i})^2}\right]\\
        &+ \sum_j \frac{\partial u_{\text{pred}}}{\partial \phi_{xj}(x_f^{i})}\left[\frac{-2jx_f^{i}}{(1-(x_f^{i})^2)^{\frac{3}{2}}}\right]\\
		&+ \sum_{m,n} \frac{\partial^2 u_{\text{pred}}}{\partial \phi_{ym}(y_f^{i})\partial \phi_{yn}(y_f^{i})}\left[\frac{4mn}{1-(y_f^{i})^2}\right]\\
        &+ \sum_m \frac{\partial u_{\text{pred}}}{\partial \phi_{ym}(y_f^{i})}\left[\frac{-2my_f^{i}}{(1-(y_f^{i})^2)^{\frac{3}{2}}}\right]\, ,
	\end{aligned}
\end{equation}
where, $j, k = 1, 2, 3, ...$ and $m, n = 1, 2, 3, ...$ are qubit indices for each feature map. Index $i = 1, 2, ..., N_f$ corresponds to collocation points.

The calculation of the first and second-order derivatives of quantum circuits concerning all qubits are described by \cref{eq:first_p_derivative} and \cref{eq:second_p_derivative}, 
\begin{equation} \label{eq:first_p_derivative}
\begin{bmatrix}
    \frac{\partial u}{\partial x} \vspace{0.1cm}\\
    \frac{\partial u}{\partial y}
\end{bmatrix} = 
    \begin{bmatrix}
    \frac{\partial u_{\text{pred}}}{\partial \phi_{xj}(x_{f}^{i})} \vspace{0.1cm}\\
    \frac{\partial u_{\text{pred}}}{\partial \phi_{ym}(y_{f}^{i})} \\
\end{bmatrix},
\end{equation}
\begin{equation} \label{eq:second_p_derivative}
\begin{split}
\begin{bmatrix}
    \frac{\partial^2 u}{\partial x^2} & \frac{\partial^2 u}{\partial x \partial y} \vspace{0.1cm}\\
    \frac{\partial^2 u}{\partial y \partial x} & \frac{\partial^2 u}{\partial y^2}
\end{bmatrix} = \hspace{1.4cm}\\ 
\begin{bmatrix}
    \frac{\partial^2 u_{\text{pred}}}{\partial \phi_{xj}(x_{f}^{i})\partial \phi_{xk}(x_{f}^{i})} &\frac{\partial^2 u_{\text{pred}}}{\partial \phi_{xj}(x_{f}^{i})\partial \phi_{yn}(y_{f}^{i})}\vspace{0.1cm}\\
    \frac{\partial^2 u_{\text{pred}}}{\partial \phi_{ym}(y_{f}^{i}) \partial \phi_{xk}(x_{f}^{i})} &\frac{\partial^2 u_{\text{pred}}}{\partial \phi_{ym}(y_{f}^{i})\partial \phi_{yn}(y_{f}^{i})}
\end{bmatrix}.
\end{split}
\end{equation}

\section{Results}\label{sec2}
For this work, we investigate and benchmark our proposed methodology on selected problems, based on earlier literature \cite{kyriienko2021solving, Kharazmi2021}. We start with a highly nontrivial dynamics problem such as the Riccati equation which is quadratic in an unknown function. Then we consider a coupled first-order differential equation, formulated as a system of DEs.  In addition, a second-order linear differential equation and the nonlinear Duffing equation are investigated. Finally, a problem on the two-dimensional Poisson’s equation presented by Kharazmi et al. \cite{Kharazmi2021} is analyzed. All simulations are performed using Pennylane software \cite{bergholm2018pennylane}.

\subsection{Complex Dynamics of Riccati Equation}
As a first example, a nonlinear differential equation of the general form known as the Riccati equation is considered, which has a rapidly oscillating non-periodic solution. This differential equation is given for a function $u(x)$ depending on the variable $x$ as an initial value problem with
\begin{equation}\label{eq:Problem1_riccati}
\begin{split}
    \frac{\td u}{\td x} - 4u + 6 u ^2 - \sin{(50 x)}
    - u \cos{(25 x)} + \frac{1}{2} &= 0, \\
	u(x=0) &= u_0
 \end{split}
\end{equation}
\begin{figure*}[ht]
	\centering
	\begin{subfigure}{0.3\textwidth}
		\includegraphics[width=\textwidth]{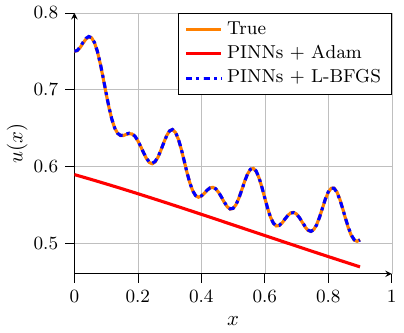}
		\caption{}
		\label{fig:Problem1a}
	\end{subfigure}
	\hfill
	\begin{subfigure}{0.3\textwidth}
		\includegraphics[width=\textwidth]{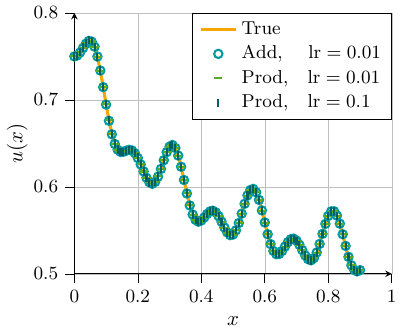}
		\caption{}
		\label{fig:Problem1b}
	\end{subfigure}
	\hfill
	\begin{subfigure}{0.3\textwidth}
		\includegraphics[width=\textwidth]{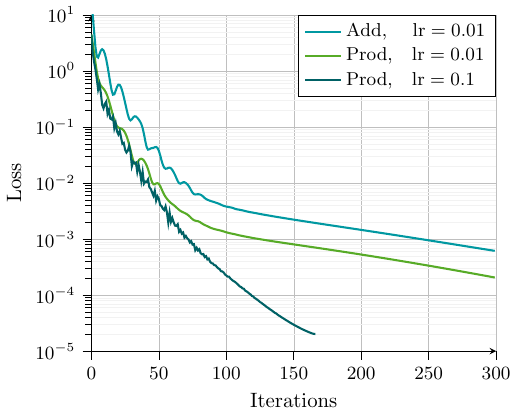}
		\caption{}
		\label{fig:Problem1c}
	\end{subfigure}
	\caption{Solution of Riccati equation \cref{eq:Problem1_riccati} and optimizer convergence. (\subref{fig:Problem1a}) Visualization of comparison between true solution obtained numerically by the Runge-Kutta method and a classical machine learning approach using PINNs with Adam and L-BFGS optimizer, (\subref{fig:Problem1b}) results for the quantum approach using the summation $\hat{\mathcal{C}} = \sum_j^N Z_j$ (referred as Add) and tensor product $\hat{\mathcal{C}} = \bigotimes_{j=1}^{N}Z_j$ (referred as Prod) of Pauli-Z operators for different learning rates (lr) and (\subref{fig:Problem1c}) their corresponding losses over the number of optimization iterations.}%
	\label{fig:Problem1}%
\end{figure*}
where the initial condition is given by $u_0 = 0.75$. The domain $x \in [0, 0.9]$ is discretized into 100 equidistant points. 

This DE is first solved by two classical methods: first, numerically by the explicit Runge-Kutta method of third order \cite{bogacki19893}, which is considered the true solution for further comparisons, and second, using PINNs, which is considered as the classical ML solution. In the case of PINNs, a two-layer feed-forward neural network with 32 hidden neurons was considered, which has a $\tanh$ activation function in each layer. To solve this problem using PINNs, the loss from initial conditions and differential equations is simply added as a multi-objective loss function \ref{sec:weighted_objective}.  The results are visualized in \cref{fig:Problem1a}. We first carried out the optimization with the Adam optimizer and found that it was unable to converge to the desired solution. However, when we carried out the optimization with the L-BFGS algorithm, the desired solution was reached.

For the quantum solution, a quantum circuit with 8 qubits and a depth of 8 blocks in the variational circuit with a circular entanglement layout (refer to \ref{sec:HEA}) has been considered. The differential equation has terms proportional to $u$ and $u^2$, which enables incorporating the initial condition into floating boundary handling (refer to \ref{sec:floating}). This enables faster convergence since a single combined loss function is used instead of a slower multi-objective optimization \cite{kyriienko2021solving}. 

A set of $N_f=100$ collocation points with $x_f^{i} \in \{1, \cdots, N_f\}$ are generated by equidistant grid points. The initial condition is given at $x_f^1$. Consider the circuit output as $u_{\text{pred}}$, and we define the modified output $u_{\text{m}}$ such that
\begin{equation}
	u_{\text{m}}(x_f^{i}) = \left(u_0 - u_{\text{pred}}(x_f^1) \right) + u_{\text{pred}}(x_f^{i}) \, , 
\end{equation}
\begin{figure}[t]
	\centering
	\begin{subfigure}{0.4\textwidth}
		\includegraphics[width=\textwidth]{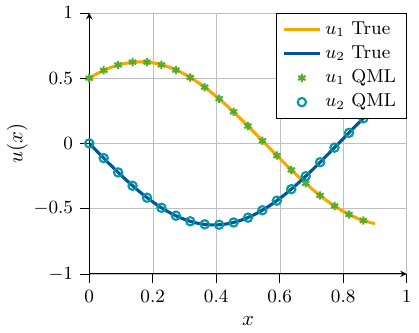}
		\caption{}
		\label{fig:Problem2a}
	\end{subfigure}
	\hspace{1cm}
	\begin{subfigure}{0.4\textwidth}
		\includegraphics[width=\textwidth]{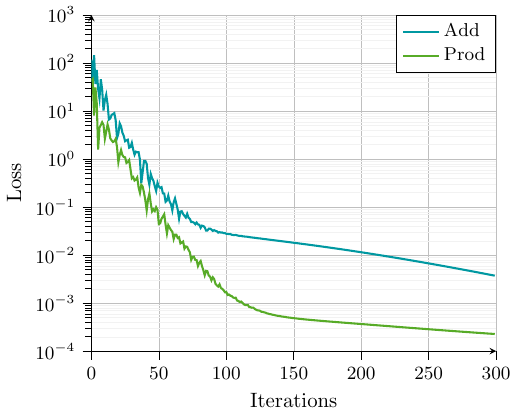}
		\caption{}
		\label{fig:Problem2b}
	\end{subfigure}
	\caption{Solving of a system of coupled differential equations (refer \cref{eq:Problem2_system1}, \cref{eq:Problem2_system2}, and \cref{eq:Problem2_system3}).  Visualization of (\subref{fig:Problem2a}) comparison between classical and quantum solution for $u_1(x)$  and $u_2(x)$ as well as (\subref{fig:Problem2b}) the loss for the summation and tensor product of Pauli-Z operators cases over the number of optimization iterations.}%
	\label{fig:Problem2}%
\end{figure}
and the residual is given by
\begin{equation}\label{eq:Problem1Residuals}
\begin{split}
    f(x_f^{i}) =& \frac{\td u_{\text{pred}}}{\td x_f^{i}} - 4u_{\text{m}}(x_f^{i}) + 6\left(u_{\text{m}}(x_f^{i})\right)^2 \\  &- \sin{(50x_f^{i})} - u_{\text{m}}\cos{(25x_f^{i})} + 1/2 \, 
\end{split}
\end{equation}
which we try to minimize. The results in \cref{fig:Problem1b} indicate that the quantum circuit can match the numerical Runge-Kutta solution as well as the solution determined classically using PINNs with L-BFGS optimizer. For all three approaches, the desired solution was obtained. The optimizer used in the quantum ML case was Adam, which was able to successfully converge to the desired solution unlike in the classical ML case where the very high precision update direction given by L-BFGS had to be used instead.

To further improve the performance of the quantum ansatz, we investigate the potential advantage of using a more expressive measurement basis to capture correlations in the circuit outcomes. In the case of using the {local observable such as} addition of Pauli-Z operators, the expectation values of each Pauli-Z operator are measured and summed up to obtain the overall result. If the quantum state is entangled, the summation of each Pauli-Z expectation value will not be able to capture the correlations between qubits. On the contrary, by {employing the global observable such as choosing the Hamiltonian as the tensor product of Pauli-Z operators}, we can capture the quantum correlations in the outcomes through entanglement in the quantum state. This modification can lead to better convergence of the cost function and improved accuracy of the solution.

In \cref{fig:Problem1c}, we compare the convergence rate towards the true solution for the same initial variational parameters of the quantum circuit, the effect of two different observables, and variation in the learning rate. We see that maintaining the same learning rate as before, the standard sum basis of Pauli-Z operators $\hat{\mathcal{C}} = \sum_j Z_j$ is outperformed by measuring instead in a product Pauli-Z basis $\hat{\mathcal{C}} = \bigotimes_{j=1}^{N}Z_j$, with errors about half an order of magnitude better. The basis further allows for the use of a much larger learning rate than before, and increasing from a rate of 0.01 to 0.1, we observe dramatically faster convergence. 

Thus, the Riccati equation is a promising first example, where the quantum approach can effectively encode the solutions, especially in systems where fast multi-partite encoding can be obtained before the measurement.

\begin{figure*}[ht]
	\centering
	\begin{subfigure}{0.25\textwidth}
		\centering
		\includegraphics[width=\textwidth]{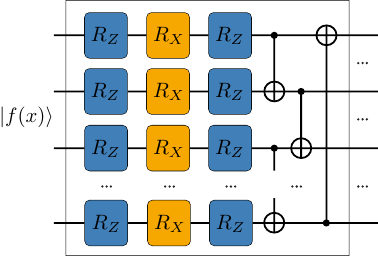}
		\caption{}
		\label{fig:Problem3a}
	\end{subfigure}
	\hfill
	\begin{subfigure}{0.35\textwidth}
		\centering
		\includegraphics[width=\textwidth]{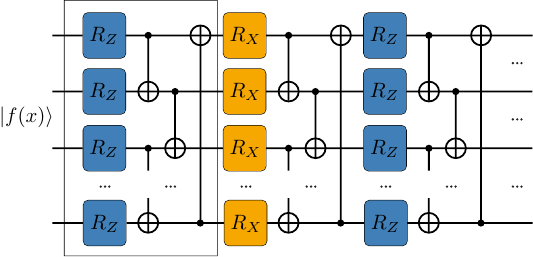}
		\caption{}
		\label{fig:Problem3b}
	\end{subfigure}
	\hfill
	\begin{subfigure}{0.3\textwidth}
		\includegraphics[width=\textwidth]{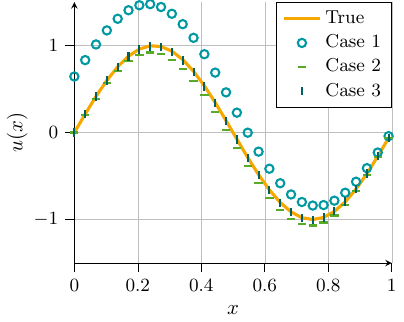}
		\caption{}
		\label{fig:Problem3c}
	\end{subfigure}
	\\ 
	\begin{subfigure}{0.3\textwidth}
		\includegraphics[width=\textwidth]{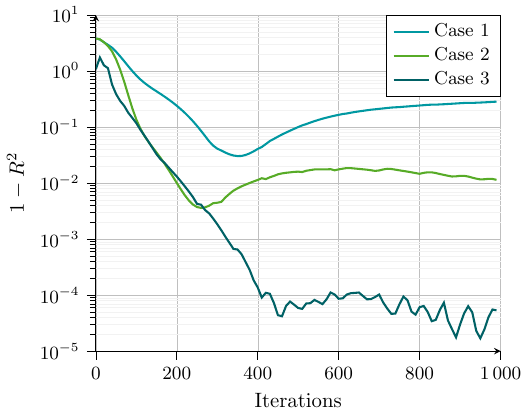}
		\caption{}
		\label{fig:Problem3d}
	\end{subfigure}
	\hfill
	\begin{subfigure}{0.3\textwidth}
		\includegraphics[width=\textwidth]{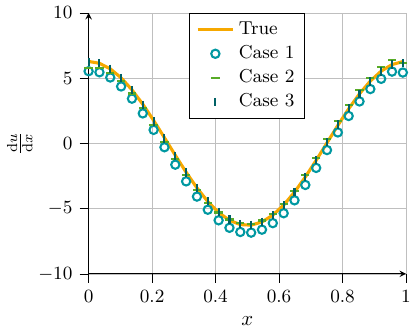}
		\caption{}
		\label{fig:Problem3e}
	\end{subfigure}
	\hfill
	\begin{subfigure}{0.3\textwidth}
		\includegraphics[width=\textwidth]{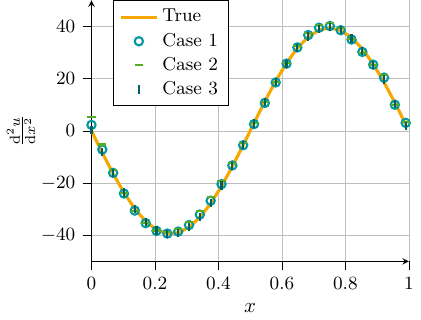}
		\caption{}
		\label{fig:Problem3f}
	\end{subfigure}
	\caption{Solution of second-order DE (refer \cref{eq:Problem3_second_order}) using constant weights for balancing multi-objective loss function. Visualization of (\subref{fig:Problem3a}) reference variational quantum circuit (refer \cref{fig:QC}) (referred to as RC). It is arranged as RZ, RX, RZ rotational layers followed by CNOT gates in cyclical order. (\subref{fig:Problem3b}) Modified variational circuit with added entangling layers in between rotational layers (referred to as AEC). (\subref{fig:Problem3c}) Solutions $u(x)$ obtained from three cases and their comparison with the true solution. Case 1 corresponds to employing of RC variational circuit, and balancing weights for multi-objective loss function $\alpha_f = \alpha_b = 1$. Case 2 corresponds to the RC variational circuit, and weights $\alpha_f = 10^{-1}$ and $\alpha_b = 10^{3}$. Case 3 corresponds to the AEC variational circuit, and weights $\alpha_f = 10^{-1}$ and $\alpha_b = 10^{3}$. (\subref{fig:Problem3d}) Comparison of accuracy in terms of $1-R^2$ measurement over iterations among three cases, and checking of (\subref{fig:Problem3e}) first derivative $\frac{du}{dx}$ and (\subref{fig:Problem3f}) second derivative $\frac{d^2u}{dx^2}$ of solution after training.}%
	\label{fig:Problem3}%
\end{figure*}

\begin{figure*}[t]
	\centering
	\begin{subfigure}{0.3\textwidth}
		\centering
		\includegraphics[width=\textwidth]{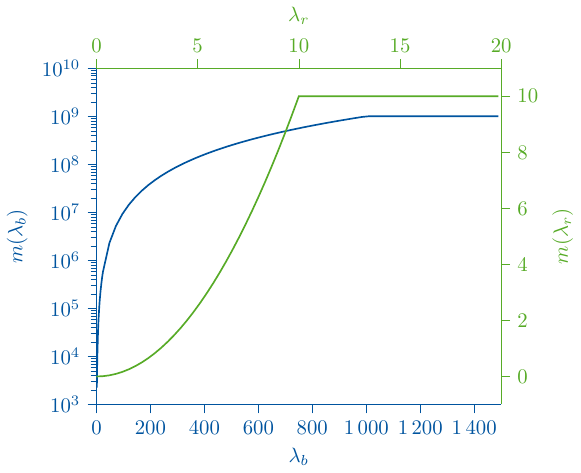}
		\caption{}
		\label{fig:Problem3p2a}
	\end{subfigure}
	\hfill
	\begin{subfigure}{0.3\textwidth}
		\centering
		\includegraphics[width=\textwidth]{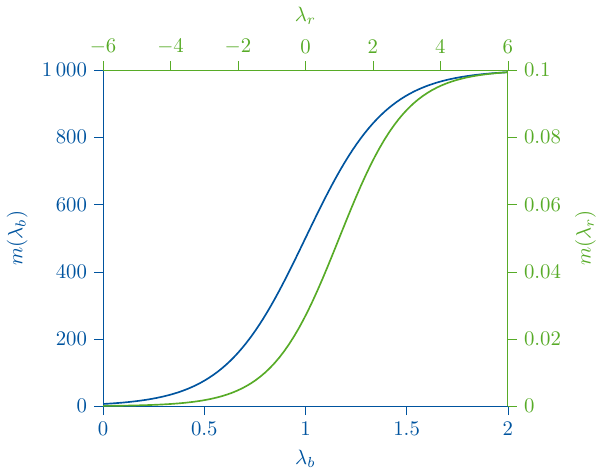}
		\caption{}
		\label{fig:Problem3p2b}
	\end{subfigure}
	\hfill
	\begin{subfigure}{0.3\textwidth}
		\includegraphics[width=\textwidth]{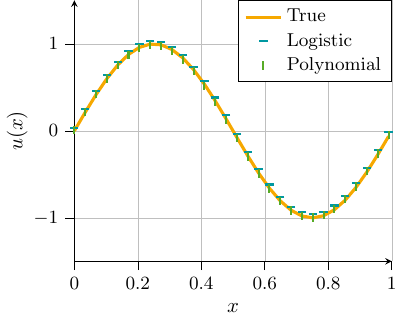}
		\caption{}
		\label{fig:Problem3p2c}
	\end{subfigure}
	\\
	\begin{subfigure}{0.3\textwidth}
		\includegraphics[width=\textwidth]{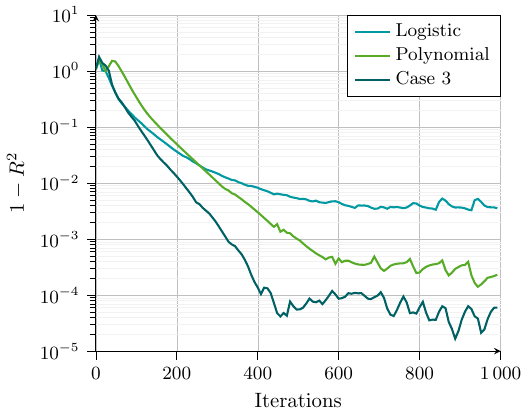}
		\caption{}
		\label{fig:Problem3p2d}
	\end{subfigure}
	\hfill
	\begin{subfigure}{0.3\textwidth}
		\includegraphics[width=\textwidth]{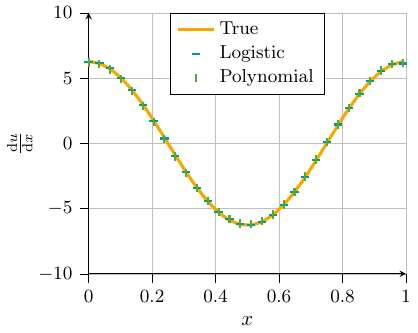}
		\caption{}
		\label{fig:Problem3p2e}
	\end{subfigure}
	\hfill
	\begin{subfigure}{0.3\textwidth}
		\includegraphics[width=\textwidth]{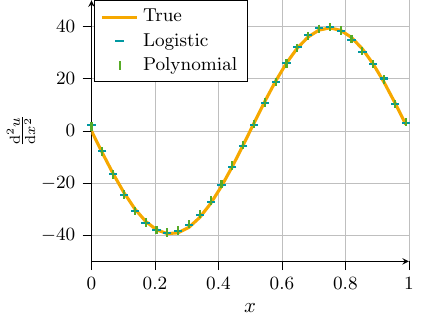}
		\caption{}
		\label{fig:Problem3p2f}
	\end{subfigure}
	\caption{Solution of second-order DE (refer \cref{eq:Problem3_second_order}) using self-adaptive weights approach. Visualization of (\subref{fig:Problem3p2a}) polynomial mask {$m(\lambda)$} for residual {$(r)$} and for boundary {$(b)$}, (\subref{fig:Problem3p2b}) logistic mask {$m(\lambda)$} for residual {$(r)$} and for boundary {$(b)$}, (\subref{fig:Problem3p2c}) solution {$u(x)$} comparison between logistic and polynomial cases, (\subref{fig:Problem3p2d}) accuracy comparison in terms of $1-R^2$ measurement over iterations for logistic and polynomial cases. In addition, the accuracy curve from case 3 is also added here for reference. Testing of (\subref{fig:Problem3p2e}) first derivative $\frac{\td u}{\td x}$ and (\subref{fig:Problem3p2f}) second derivative $\frac{\td^2u}{\td x^2}$ of predicted solution $u(x)$.} %
	\label{fig:SAPINNs}%
\end{figure*}
\subsection{System of Differential Equations}
We now turn our methodology to another problem, first proposed in Kyriienko et al. \cite{kyriienko2021solving}, where a first-order system of differential equations is investigated. In this case, a vector $\boldsymbol{u}(x) = \begin{bmatrix} u_1(x), u_2(x)\end{bmatrix}$ describes two modes, whose progression in time is $x$ given by
\begin{align}
	F_1\left(\td_x\boldsymbol{u}, \boldsymbol{u}, x\right) =& \frac{\td u_1}{\td x} - a_1 u_2 - a_2 u_1 = 0\, ,\label{eq:Problem2_system1}\\
	F_2[\td_x\boldsymbol{u}, \boldsymbol{u}, x] =& \frac{\td u_2}{\td x} + a_2 u_2 + a_1 u_1 = 0\, , \label{eq:Problem2_system2}\\
    \boldsymbol{u}(0) =& \begin{bmatrix} u_1(0), u_2(0)\end{bmatrix} = \begin{bmatrix} u_{1,0}, u_{2,0}  \end{bmatrix} \, \label{eq:Problem2_system3},
\end{align}
where $a_1$ and $a_2$ are coupling parameters and $u_{1,0}$ as well as $u_{2,0}$ are initial conditions. The classical solution for this problem is obtained through ODE solvers by \cite{hindmarsh1983odepack, petzold1983automatic}. To achieve a quantum solution, a quantum circuit comprising three qubits and a depth of seven blocks in the variational circuit is considered (see \ref{sec:HEA}). An Adam optimizer is once again chosen, with a learning rate of 0.1. The observables are compared for the cases of addition and product as explained in the previous example. The domain is considered to be $x \in [0, 0.9]$, discretized into 100 equidistant grid points. The parameters for this problem are $a_1 = 5, \, a_2 = 3$ and initial conditions are $u_{1,0}=0.5$ and  $u_{2,0}=0$. 
\begin{figure*}[t]
	\centering
	\begin{subfigure}{0.235\textwidth}
		\centering
		\includegraphics[width=\textwidth]{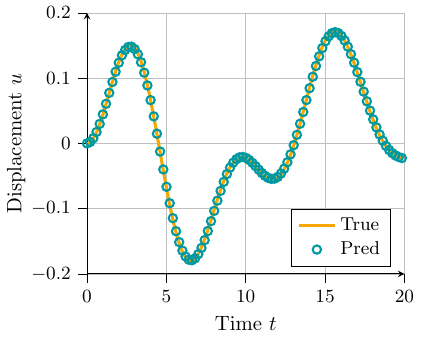}
		\caption{}
		\label{fig:Problem5a}
	\end{subfigure}
	\begin{subfigure}{0.235\textwidth}
		\centering
		\includegraphics[width=\textwidth]{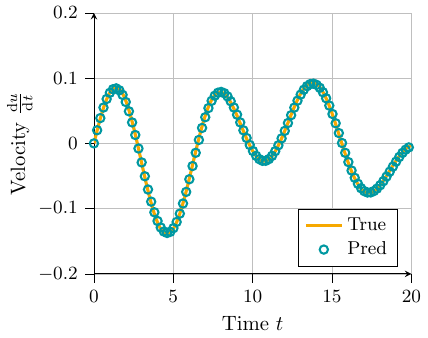}
		\caption{}
		\label{fig:Problem5b}
	\end{subfigure}
	\begin{subfigure}{0.28\textwidth}
		\includegraphics[width=\textwidth]{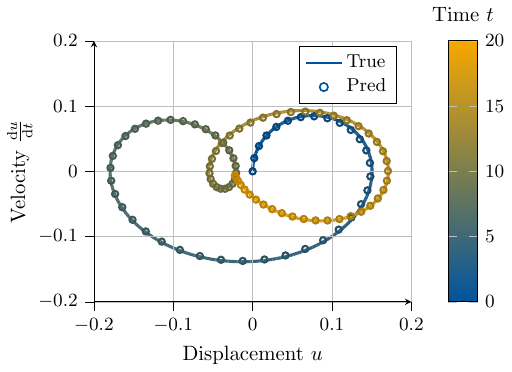}
		\caption{}
		\label{fig:Problem5c}
	\end{subfigure}
	\begin{subfigure}{0.23\textwidth}
		\includegraphics[width=\textwidth]{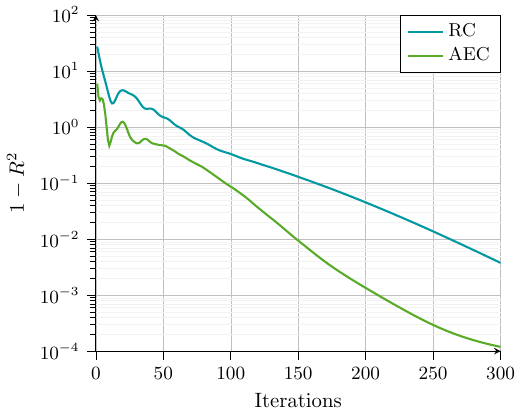}
		\caption{}
		\label{fig:Problem5d}
	\end{subfigure}
	\caption{Solution of Duffing equation (refer \cref{eq:Duffing}). Visualization of (\subref{fig:Problem5a}) displacement $(u)$ vs. time $(t)$, (\subref{fig:Problem5b}) velocity $(\frac{\td u}{\td t})$ vs. time $(t)$, (\subref{fig:Problem5c}) velocity vs. displacement over the time evolution, also known as the phase-space plot of Duffing oscillation. The color bar on the right depicts the time $(t)$. The accuracy comparison (\subref{fig:Problem5d}) in terms of $1-R^2$ measurement over iterations for reference (RC) and added entanglement circuit (AEC).}
	\label{fig:Problem5}%
\end{figure*}
The results obtained from global observable are shown in \cref{fig:Problem2a}. It is clear that the proposed quantum algorithm accurately captures the known solution. In \cref{fig:Problem2b}, we again compare the modified observables such as the tensor product with the summation of Pauli-Z operators. Once again we see a large impact in the accuracy and convergence towards the known solution.

\subsection{Second-Order Linear Differential Equation}
\label{subsec:SecondDE}
As a further generalization, we now consider a second-order DE given for the function $u(x)$ with
\begin{equation}\label{eq:Problem3_second_order}
    \frac{\td^2u}{\td x^2} + 4\pi^2\sin{(2\pi x)} = 0,
\end{equation}
where $u(0)=0$ and $u(0.99) = \sin{2\pi 0.99}$. The analytical solution is given with $u(x) = \sin{(2\pi x)}$. Note that the Dirichlet boundary conditions at the right end are evaluated at $x=0.99$ instead of $x=1.0$ due to the condition explained in \cref{sec:second_derivative_section}. A set of $N_{f}=30$ collocation points with $x_f^{i} \in \{1, \dots , N_{f}\}$ are generated by equidistant grid points. Using the Chebyshev feature map 
\begin{equation}
    \phi_j(x_{f}^{i}) = 2j \arccos{(x_{f}^{i})},
\end{equation}
the residual $f(x_{f}^{i})$ for a single data point is given for the circuit output $u_{\text{pred}}$ by 
\begin{equation}\label{eq:residual_second_order}
	f(x_{f}^{i}) = \frac{\td^2 u_{\text{pred}}}{\td (x_f^{i})^2} + 4\pi^2 \sin{(2 \pi x_{f}^{i})} \, ,
\end{equation}
where the second derivative $\frac{\td^2 u_{\text{pred}}}{\td (x_f^{i})^2}$ can be determined using the chain rule \cref{eq:d2u_dx2_derivation}.

The differential equation in this problem does not contain the terms $u(x)$ in it. Therefore, the Dirichlet boundary conditions cannot be applied using floating boundary handling (refer \ref{sec:floating}). Hence, the loss function is formulated as a multi-objective loss function (\ref{sec:weighted_objective}). This loss function requires balancing the loss terms $\alpha_f, \alpha_b$. Therefore, two cases are compared: in the first case, we show the effect of unbalanced loss terms, where the weights of the total loss function are given with $\alpha_f = \alpha_b = 1$ (Case 1), and in the second case, the weights are constant factors and they are heuristically chosen as $\alpha_f = 10^{-1}$ and $\alpha_b = 10^{3}$ (Case 2). For these cases, a circuit with three qubits and a depth of seven blocks in the variational circuit \ref{sec:HEA} has been considered where the observables are chosen with the standard $\hat{\mathcal{C}} = \sum_j Z_j$.

The obtained solution for cases 1 and 2 together with the first and second derivative of the solution is visualized in \cref{fig:Problem3c}, \cref{fig:Problem3e}, and \cref{fig:Problem3f}. The accuracy of the solutions is compared through the $R^{2}$ score using the term $1-R^2$ (see \cref{fig:Problem3d}). The results comparison between cases 1 and 2 suggest that balancing weights are important while handling multi-objective loss function. However, the accuracy of the quantum ML circuit has trouble in stabilizing below 1\%.

To further improve the accuracy of the given problem, we study the effect of more entangling layers in the variational circuit. Consider the variational circuit represented in \cref{fig:Problem3a} as the reference circuit (RC) corresponding to variational part of \cref{fig:QC}. A modified variational circuit with more entangling layers is represented in \cref{fig:Problem3b}, which we refer to as an added entanglement circuit (AEC). With this modified circuit, we carry out another simulation use case (Case 3) with balanced weights of $\alpha_f = 10^{-1}$ and $\alpha_b = 10^{3}$. For all cases and as in the previous differential equations, an Adam optimizer was employed, where now the learning rate was specified at $0.01$ for a total of $1000$ iterations. 

In the comparison between cases 2 and 3, the results suggest that adding entangling layers has enhanced the reachability of the circuit. It can improve the convergence and accuracy with the same number of variational parameters and their initial values. Therefore, entangling layers have played a key role in improving the accuracy while solving the second-order differential equation. While this worked for this example, \cite{mcclean2018barren} indicates that added entanglement hinders the optimization leading to a barren plateau. Hence, care must be taken while using larger circuits and emerging methods must be further explored to mitigate this issue \cite{kulshrestha2022beinit, larocca2022diagnosing, mele2022avoiding}.

To further explore the generalized case of variable balancing weights of the multi-objective loss function, an alternative approach based on SAPINNs \cite{mcclenny2023self} is incorporated (refer \cref{section:SAPINNs}). 
Based on the specifications provided in \cref{sec:mask_functions}, we define the mask functions for this problem as shown in \cref{fig:Problem3p2a} and \cref{fig:Problem3p2b}.

We consider the modified variational circuit with added entangling layers (see \cref{fig:Problem3b}) and the same initial variational parameters from cases 1, 2, and 3 for a fair comparison. From the displacement, first, and second derivative graphs, both the poly and sigmoid masks can approximate the solution to a satisfactory level (see \cref{fig:Problem3p2c}, \cref{fig:Problem3p2e} and \cref{fig:Problem3p2f}). The accuracy comparison in \cref{fig:Problem3p2d} shows that the simple constant weights (Case 3) were able to perform better. However, the results prove that the approach of self-adaptive weights is suitable for variational quantum circuits and can be used for various problems. Furthermore, as the complexity of the problem increases with a complex optimization landscape and more loss terms in the multi-objective loss function, the approach of self-adaptive weights will provide us with better hyperparameters to tweak as compared to repetitive trials with non-adaptive constant weights. Therefore, this approach might have more potential to save time in exploring hyperparameters and have a better convergence as compared to non-adaptive constant weights.

\subsection{Second-Order Nonlinear Differential Equation}\label{sec:duffing}
For the next problem, we consider a Duffing equation which is a nonlinear second-order differential equation given by,
\begin{equation}\label{eq:Duffing}
\frac{\td^2 u}{\td t^2} + \delta \frac{\td u}{\td t} + \alpha u + \beta u^3 = \gamma \cos{(\omega t)}.
\end{equation}
In this work, we have considered an example of this problem from \cite{salas2022elementary}, where the parameters are given by, $\alpha = 1, \beta = 1, \delta = 0.1, \gamma = 0.1, \omega = 0.4$. The initial conditions are given by $u(0) = 0$ and $\frac{\td u(0)}{\td t} = 0$. The time evolution to solve the problem is chosen from $t=0$ to $t=20$. As a reference solution (also referred to as true solution), we employed an explicit Runge-Kutta method of fifth order \cite{dormand1980family}. For the quantum solution, the dependent variable is suggested to be not close to 1.0 due to singularity (refer \cref{sec:second_derivative_section}). Therefore, the dependent variable in the Duffing equation is scaled and solved within the interval $t=[0, l]$, where we choose the limit, $l=0.9$. Consider the total time of the evolution as $T=20$ and the scaled time variable as $\tau = \frac{t}{T}l$. Then the modified differential equation is given by,
\begin{equation}
    \left(\frac{l}{T}\right)^2 \frac{\td^2 u}{\td \tau^2} + \frac{l}{T} \delta \frac{\td u}{\td \tau} + \alpha u + \beta u^3 = \gamma \cos{\left(\frac{\omega T \tau}{l}\right)}.
\end{equation}
The solution grid is chosen to be the discretization of 50 equidistant time intervals. Similar to the previous problem, the cost function is formulated as $C=\alpha_f \text{MSE}_f + \alpha_b \text{MSE}_b$ (see \ref{sec:weighted_objective}), where the weights $\alpha_f = 1$ and $\alpha_b = 1$. In this case, there was no need for balancing these two terms while their magnitude was found to be of similar order. The hyperparameters chosen for this problem are five qubits and the depth of eight blocks in the variational part of the quantum circuit (refer to \ref{sec:HEA}). The observables are chosen to be $\hat{\mathcal{C}} = \sum_j Z_j$. To test the applicability of the proposed added entangling layers, we have considered two cases similar to the previous problem. The first case is where we chose the reference circuit \cref{fig:Problem3a} and the second one is the added entangling circuit \cref{fig:Problem3b}. For both cases, the initial variational parameters are the same for fair comparison. An Adam optimizer with a learning rate of 0.01 is employed with 300 iterations. After solving the problem at coarse 50 discretized points, the trained circuit is tested to predict the solution at 500 equidistant time steps for visualization of a smooth solution. This is one of the advantages of the spectral methods over finite difference methods, namely of having smooth interpolation. The results presented in the \cref{fig:Problem5a} and \cref{fig:Problem5b} show that the displacement and velocity are well-matched with the Runge-Kutta solution. We have also depicted the phase space plot in \cref{fig:Problem5c}, which is velocity vs. displacement, to visualize the resulting Duffing oscillation. Furthermore, the accuracy plot using the measurement $1-R^2$ (see \cref{fig:Problem5d}) clearly shows that the added entangling circuit has better performance as compared to the reference circuit.
\begin{figure*}[t]
	\centering
	\begin{subfigure}{0.35\textwidth}
		\includegraphics[width=\textwidth]{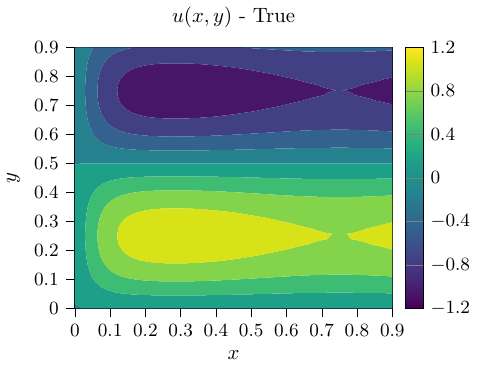}
		\caption{}
		\label{fig:Problem4a_b}
	\end{subfigure}
	\begin{subfigure}{0.35\textwidth}
		\includegraphics[width=\textwidth]{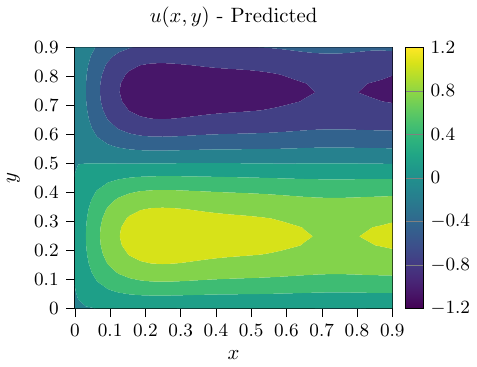}
		\caption{}
		\label{fig:Problem4a_c}
	\end{subfigure} \vspace{0.1cm}
        \\
	\begin{subfigure}{0.3\textwidth}
		\includegraphics[width=\textwidth]{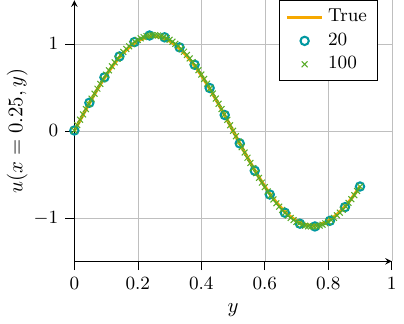}
		\caption{}
		\label{fig:Problem4a_d}
	\end{subfigure}
	\hfill
	\begin{subfigure}{0.3\textwidth}
		\includegraphics[width=\textwidth]{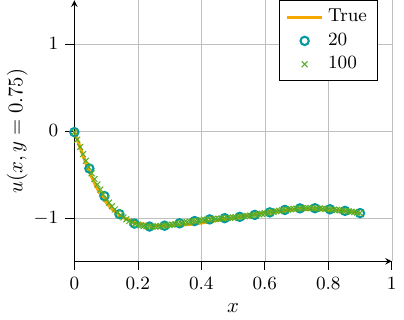}
		\caption{}
		\label{fig:Problem4a_e}
	\end{subfigure}
	\hfill
	\begin{subfigure}{0.3\textwidth}
		\includegraphics[width=\textwidth]{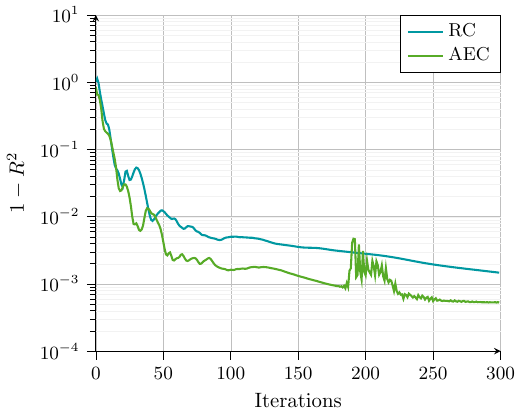}
		\caption{}
		\label{fig:Problem4a_f}
	\end{subfigure}
	\caption{Solution of 2D Poisson's equation (refer \cref{eq:Poisson2D}). Visualization of (\subref{fig:Problem4a_b}) true and (\subref{fig:Problem4a_c}) predicted solution $u(x, y)$ of Poisson's equation with color bar mapping the function values $u(x, y)$ to contour plot. Interpolation test for 20 and 100 equidistant points at (\subref{fig:Problem4a_d}) $x=0.25$ and (\subref{fig:Problem4a_e}) $y=0.75$, where the problem is originally solved at 30 equidistant points. \subref{fig:Problem4a_f}) The accuracy comparison in terms of $1-R^2$ measurement over iterations for reference (RC) and added entanglement circuit (AEC).}%
	\label{fig:Problem4a}%
\end{figure*}
\subsection{Second-Order Partial Differential Equation}\label{sec:poisson_results}
As a final use-case, in the domain of PDEs, we consider a two-dimensional Poisson's equation \cite{Kharazmi2021} given by
\begin{equation}
\begin{aligned}
    \Delta u(x, y) = g(x, y) \; \text{with} \; x, y \in [0, 0.9] \times [0, 0.9] \, ,
	\label{eq:Poisson2D}
 \end{aligned}
\end{equation}
where the homogenous Dirichlet boundary conditions and the right-hand side $g(x,y)$ are chosen in such a way that the exact solution is given by
\begin{equation}
	u_{\text{true}}(x, y) = (0.1 \sin{(2\pi x)} + \tanh{(10x)})\sin{(2\pi y)} \, .
\end{equation}

The quantum formulation of this problem is described in detail in \cref{sec:methods_Poisson}. The cost function for this problem consists of residual loss and boundary conditions. For calculating the residual loss, the domain for this problem is discretized into $30 \times 30$ equidistant points totaling 900 collocation points $N_f(x_f, y_f)$. The solution on the boundary is represented by $N_b$ labeled data points $\{x_b^{i}, y_b^{i}, u_b^{i}\}$ and the Dirichlet boundary conditions are given by $u_b(x, 0), u_b(x, 0.9), u_b(0, y),$ and $u_b(0.9, y)$. Note that, we are examining a slightly altered problem domain than the usual boundary points $x=1, y=1$ due to computational costs and numerical instabilities in the differential operator leading to a modified boundary (see \cref{sec:second_derivative_section}). Finally, the cost function is formulated as $C=\alpha_f \text{MSE}_f + \alpha_b \text{MSE}_b$ (see \ref{sec:weighted_objective}), where the weights $\alpha_f = 10^{-1}$ and $\alpha_b = 10^3$. Adam optimizer with a learning rate of $0.1$ is employed with 300 iterations.

The circuit architecture for this problem is chosen to be 3 qubits for variable $x$ and 3 qubits for variable $y$ with a depth of 7 blocks in the variational circuit (see \ref{sec:HEA}). Two cases are considered for this problem for variational circuits, first RC and second AEC to study the applicability of added entangling layers. The predicted solution \cref{fig:Problem4a_c} from the AEC variational circuit is represented and compared with the true solution \cref{fig:Problem4a_b}. This approach has distinctly better convergence and accuracy when compared to the standard RC variational circuit, as illustrated in \cref{fig:Problem4a_f}. 

To further test the generalization performance of the trained circuit on the unseen data points, we have carried out an interpolation test within the solved domain. The domain is initially discretized to 30 equidistant points in between boundaries. Therefore, the interpolation test is conducted on a coarse grid (large steps) of 20 equidistant points and a fine grid (small steps) of 100 equidistant points. This test is performed on specific critical lines, such as $x=0.25$ (see \cref{fig:Problem4a_d}) and $y=0.75$ (see \cref{fig:Problem4a_e}). Based on these results, it can be observed that the trained circuit can interpolate and accurately predict the solution in unseen data points. 

\section{Discussion}
Physics-informed quantum machine learning has an enormous potential for solving a diverse spectrum of problems related to differential equations. A variety of approaches have been presented to solve DEs starting with the HHL algorithm, nonlinear variational solvers, quantum kernel methods, as well as using feature map encoding. In this work, the latter method is successfully combined with a self-adaptive variational quantum neural network to tackle challenging differential equations. For initial value problems handled with floating boundary conditions, the proposed change in the measurement observables of the quantum circuit from the summation to the tensor product of Pauli-Z operators results in a remarkable accuracy improvement. The advantage of adding more entangling layers in the variational circuit is witnessed as an improvement in the accuracy while solving second-order differential equations.

Initially, a highly nontrivial Riccati equation is solved, and a system of differential equations is investigated comparing the results to existing solutions. In addition, a second-order linear differential equation is analyzed. The importance of balancing the multi-objective loss function is presented using constant balancing weights. To further explore the generalized case, an alternative approach based on self-adaptive weights is presented incorporating the optimization of balancing weights. Furthermore, a complex example of the nonlinear Duffing equation is analyzed. Lastly, a second-order 2D partial differential equation is investigated. Therefore, a new quantum circuit is designed that can incorporate multiple independent variables.

Overall, the modified quantum circuits with added entangling layers lead to more accurate results in fewer epochs while solving second-order ordinary and partial differential equations. The enhancement in performance can be attributed to the fact that quantum circuits containing a larger number of entangling layers will result in a higher degree of expressibility. Consequently, the quantum states within the Hilbert space will be more thoroughly explored, leading to more precise outcomes. This work showcases the promise of quantum machine learning in addressing a variety of differential equations, offering an innovative and efficient solution approach compared to classical methods. The adaptability of the proposed framework to different problem domains and its potential for further extensions, such as 3D partial differential equations and diverse applications e.g. in engineering or materials science, underscore the significance of advancing the field of quantum computing for solving real-world problems. Nevertheless, many questions remain regarding the capabilities of the presented work when applied to different problems such as nonlinear mechanics e.g. modeling finite-strain plasticity, solving the Navier-Stokes equation, or even coupled problems for fluid-structure interaction. A further question that still needs to be addressed is whether a potential quantum advantage can be achieved. Thus, the presented work is an initial starting point for further applications of QML presenting promising results to approach these problems.

\section*{Author Contribution}
AS was responsible for conceptualization, methodology, implementation, original draft writing, and editing. RA and FM gave guidance and are responsible for editing and reviewing the draft.

\section*{Data Availability}
The datasets used and/or analyzed during the current study are available from the corresponding author upon reasonable request.

\section*{Acknowledgements}
We would like to thank Mikhail Itskov and Jos\'e Jesus for their helpful discussion and support. FM and AS received funding through the European Union’s Horizon Programme (HORIZONCL4-2021-DIGITALEMERGING-02-10), Grant Agreement 101080085 (QCFD).

\bibliography{bibliography.bib}

\newpage
\clearpage
\onecolumn

\appendix

\begin{center}
    \section*{Appendix}
\end{center}

\section{Weighted Multi-Objective Loss Function}\label{sec:weighted_objective}
The differential equations solved using variational quantum algorithms incorporating Chebyshev feature maps require minimizing the cost function which consists of the residual term for the differential equation itself and the boundary conditions. 

Firstly, if the residual term for the differential equation is given by $f(x_f^{i})$, the mean squared error $\text{MSE}_f$ can be determined using $N_f$ collocation points by 
\begin{align}
	\text{MSE}_f = \frac{1}{N_f}\sum_{i=1}^{N_f}\left( f(x_f^{i}) \right)^2 \, .
	\label{eq:MSER}
\end{align}
Secondly, the boundary loss term $\text{MSE}_b$, which is only evaluated for all boundary points $N_b$ is given by,
\begin{align}
	\text{MSE}_b = \frac{1}{N_b}\sum_{i = 1}^{N_b} \left(u_{\text{pred}}(x_b^{i}) - u_b(x_b^{i})\right)^2 \, .
	\label{eq:MSEBC}
\end{align}
While solving the differential equation, it is required to minimize the total loss function by combining both residual and boundary loss terms. However, due to differences in the order of magnitude of these terms, it is beneficial to balance the loss terms in magnitude while optimizing. Therefore, a weighted approach for the total loss $C$ can be formulated using
\begin{align}
	C = \alpha_f \text{MSE}_f + \alpha_b \text{MSE}_b \, ,
	\label{eq:TotalLoss}
\end{align}
where $\alpha_f$ and $\alpha_b$ are weights counteracting the imbalance of order between the residual and boundary loss terms. 

\section{Solving Differential Equations Using Quantum Chebyshev Feature Maps} \label{sec:QFM}
For the quantum circuit shown in the \cref{fig:QC}, this section details the methodology employed in this work. As Kyriienko et. al \cite{kyriienko2021solving} stated, the objective is to approximate the solutions of the differential equations as quantum circuits parameterized by a variable $x \in \mathbb{R}$ (or a collection of $\nu$ variables, $\mathbf{x} \in \mathbb{R}^{\nu}$). For brevity, a simplified single variable notation $x$ is used for the generalized case of $\nu$ variables. The data is encoded into the quantum circuit using a \textit{quantum feature map} $\hat{\mathcal{U}}_{\phi}(x)$, where \textit{nonlinear} function of variables $\phi(x)$ are predefined. These quantum feature maps encode the data as amplitudes of the quantum state $\hat{\mathcal{U}}_{\phi}(x) |0 \rangle$ from some initial state like $|0 \rangle$. \\
Compared to other encoding techniques such as amplitude encoding which require access to the amplitudes, a quantum feature map encoding has several advantages. It can represent latent space encoding (multi-dimensional representation of compressed data) and it is controlled by gate parameters, mapping real parameter $x$ to the corresponding variable value. Upon encoding, a vector of variational parameters $\theta$ are added by a variational quantum circuit $\hat{\mathcal{U}}_{\theta}$. These parameters are updated by a quantum-classical optimization loop, similar to classical machine learning. \\
The resulting state $|f_{\phi, \theta} (x)\rangle = \hat{\mathcal{U}}_{\theta}\hat{\mathcal{U}}_{\phi}(x) |0 \rangle$  with optimized parameters contains the $x$-dependent amplitudes driven to represent the desired function. However, the resulting state is a quantum state and the desired function $f(x)$ is a classical function. Therefore, the real-valued function $f(x)$ can be brought out by calculating the expectation value of a predefined Hermitian cost operator $\hat{\mathcal{C}}$, such that 
\begin{equation}
    f(x) = \langle f_{\phi , \theta}(x) | \hat{\mathcal{C}} | f_{\phi, \theta}(x)\rangle.
\end{equation}
Since solving differential equations requires computing gradients $\frac{df(x)}{dx}$, the quantum way of achieving this is by differentiating the quantum feature map circuit with
\begin{equation}\label{eq:derivative}
	\frac{d\hat{\mathcal{U}}_{\phi}(x)}{dx} = \sum_j \hat{\mathcal{U}}_{d\phi ,j}(x),		
\end{equation}
where the index $j$ runs through the individual quantum operations used in the feature map encoding. Here, derivatives are represented by the product derivative rule. When a quantum feature map is written by strings of Pauli matrices or any involutory matrix, the \textit{parameter shift rule} can be used. Then the function derivatives are written as a sum of expectations with
\begin{equation}\label{eq:derivative_2}
	\frac{df(x)}{dx} = \frac{1}{2} \sum_j \left( \langle f^+_{d\phi ,j, \theta}(x)|\hat{\mathcal{C}} |f^+_{d\phi ,j, \theta}(x) \rangle -  \langle f^-_{d\phi ,j, \theta}(x)|\hat{\mathcal{C}} |f^-_{d\phi ,j, \theta}(x) \rangle \right) \, ,
\end{equation}
where $|f^{\pm}_{\phi ,j, \theta} (x)\rangle$ is defined through the parameter shifting. The second-order derivative $\frac{d^{2} f(x)}{dx^{2}}$ can be determined using four shifted terms for each generator. For simulators, automatic differentiation (AD) comes as handy in this situation. AD represents the function derivatives by an exact analytical formula using a set of simple computational rules, as opposed to numerical differentiation. \\
The solution of differential equations is represented by constructing and defining the conditions for the quantum circuit, such that
\begin{equation}
	F[\{d^m f_n/dx^m\}_{m,n}, \{f_n(x)\}_n, x] = 0,
\end{equation}
where the functional $F$ is provided by the problem for function derivatives of different order $m$ and function/variable polynomials of varying degree $n$. For brevity, consider $f$ as both a function and a vector of functions. Then this functional can be given as a task for optimization problem with loss function $\mathcal{L}_{\theta}[d_xf, f, x]$. In other words, this is a minimization problem $F[x]|_{x \rightarrow x_i}$ at points in the set $\mathbb{X} = \{x_i\}_{i=1}^M$, as well as taking boundary conditions into account. After the optimization, the updated angles
\begin{equation}
	\theta_{\text{opt}} = \underset{\theta}{\text{argmin}}(\mathcal{L}_{\theta}[d_xf, f, x])
\end{equation}
are used to represent the solution 
\begin{equation}
	f(x)|_{\theta \rightarrow \theta_{\text{opt}}} \approx f(x)_{\text{desired}}.
\end{equation}

In the following subsections, we discuss in detail the Chebyshev feature maps, variational quantum circuits, and so on. 

\subsection{Chebyshev Feature Maps}
A nonlinear quantum feature map, namely, $Chebyshev$ $feature$ $map$ changes the basis set of function representation by leveraging Chebyshev polynomials. They are defined as a single qubit rotation $R_{Y,j}(\phi [x])$, but with nonlinear function $\phi (x) = 2n \arccos{x}, n=0,1,2,\cdots$, such that the encoding circuit is written as,
\begin{equation}
	\hat{\mathcal{U}}_{\phi}(x) = \bigotimes_{j=1}^{N} R_{Y,j}(2n[j]\arccos{x}) \, .
	\label{eq:5.14}
\end{equation}
Here, the coefficient $n[j]$ depends on the qubit position $j$. Using Euler's formula, the \cref{eq:5.14} can be rewritten as,
\begin{equation} \label{eq:5.15}
    \begin{split}
        R_{Y,j}(\phi [x]) &= \exp{\left( -i \frac{2n[j] \arccos{(x)}}{2} Y_j\right)}, \\
        &= \cos{(n[j] \arccos{(x)})} I_j - i \sin{(n[j] \arccos{(x)})} Y_j.
    \end{split}
\end{equation}

To rewrite the above \cref{eq:5.15}, consider the definition of Chebyshev polynomials \cite{mason2002chebyshev}. They are determined by two sequences relating to cosine and sine functions, noted as $T_n(x)$ and $U_n(x)$. 

Chebyshev polynomials of the first kind $T_n$ are defined by,
\begin{equation}
    T_n(\cos{\theta}) = \cos{(n \theta)}.
\end{equation}
Similarly, Chebyshev polynomials of the second kind $U_n$ are defined by,
\begin{equation}
    U_n(\cos{\theta}) \sin{\theta} = \sin{((n+1)\theta)}.
\end{equation}
Using the above relations, the \cref{eq:5.15} is further decomposed into unitary operation with matrix elements defined by degree-$n$ Chebyshev polynomials of the first and second kind, such that,
\begin{equation}
    R_{Y,j}(\phi [x]) = T_n(x) I_j + \sqrt{1-x^2}U_{n-1}(x) X_j Z_j.
\end{equation}

In this work, a specific function $n[j] = j$ is chosen as it will expand the basis functions to the order of $n^{\text{th}}$ degree polynomial \cite{kyriienko2021solving}. Thus, the updated feature map is given by 
\begin{equation}
    \hat{\mathcal{U}}_{\phi}(x) = \bigotimes_{j=1}^N R_{Y,j}(2j \arccos{x}),
\end{equation}
where the encoded variable expands with the number of qubits, making a tower-like structure of polynomials with increasing $n=j$. When the polynomials multiply with each other and morph between two kinds and their degrees, the basis set will become massive. This solves the problem of large expressibility without increasing the system size and number of rotations. 

\begin{figure}[H]
\centering
\includegraphics[]{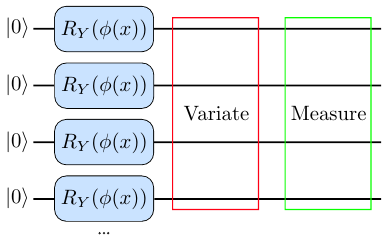}
\captionsetup{width=\textwidth}
\caption[Feature map circuit.]{Quantum Chebyshev feature map with single qubit rotations act at each qubit individually and are parameterized by a nonlinear function $\phi(x)$ of variable $x$. The variational circuit and measurement of the circuit are given by red and green blocks, respectively.}
\label{fig:Appendix_Image_1}
\end{figure}

\begin{figure}[H]
\centering
\resizebox{\textwidth}{!}{\includegraphics[]{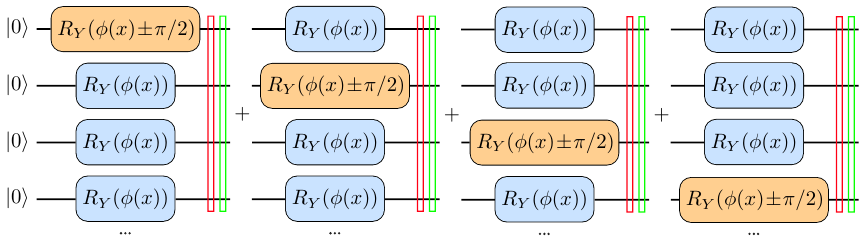}}
\captionsetup{width=\textwidth}
\caption[Derivative of the quantum feature map.]{Derivative of the quantum feature map. Differentiation with respect to $x$ is followed by the chain rule, with the expectation value of the derivative written as a sum of separate expectations with shifted phases, repeated for each $x$-dependent rotation.}
\label{fig:Appendix_Image_2}
\end{figure}

Consider the feature map circuit as in \cref{fig:Appendix_Image_1}. The derivative of the quantum circuit is calculated using the parameter shift rule (refer \cref{eq:derivative} and \cref{eq:derivative_2}) as shown in \cref{fig:Appendix_Image_2}.

\subsection{Variational Quantum Circuit}\label{sec:HEA}
Solving differential equations as a quantum circuit means converging both derivatives and functions to the desired form. This requires manipulating the latent space basis function through the variational circuit $\hat{\mathcal{U}}_{\theta}$, which is referred to as a \textit{variational quantum ansatz}. In this work, a well-known architecture, namely, \textbf{hardware efficient ansatz} is implemented.

\begin{figure}[H]
    \centering
    \includegraphics[]{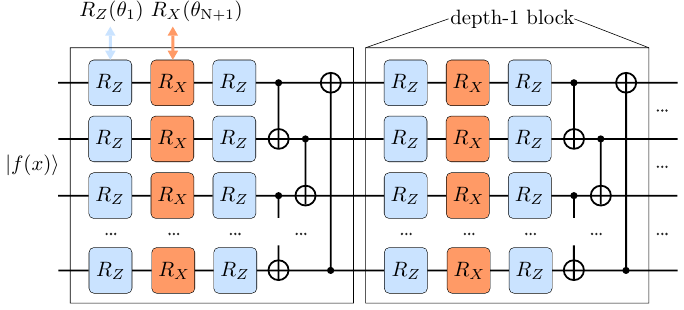}
    \captionsetup{width=\textwidth}
    \caption[Hardware efficient ansatz as a variational quantum circuit.]{Hardware efficient ansatz as a variational quantum circuit along with the representation of a single block/layer.}
    \label{fig:Appendix_Image_3}
\end{figure}

Hardware efficient ansatz (HEA) consists of layers of parameterized rotations, followed by layers of CNOT operations shown in \cref{fig:Appendix_Image_3}. A sequence of three layers of rotations $R_Z-R_X-R_Z$ are concatenated and parameterized by independent angles $\theta$ such that arbitrary single-qubit operations can be reproduced. CNOT gates act as an entangling layer. The rotations and CNOT gates together are referred to as blocks and this block is repeated for a depth of $d$ times. The number of layers $d$ is directly proportional to the circuit's expressive power (ability to represent arbitrary $N$-qubit unitary gates). However, as the parameters increase, the optimization will suffer with a problem known as \textbf{barren plateau}. The choice of ansatz is still an open question for improving trainability and requires further studies in this direction.

\subsection{Observables}
To get the output from quantum circuits, a Hermitian observable $\hat{\mathcal{C}}$ is used to measure the expectation. The scalar function $f(x)$ is then computed by expectation of $\hat{\mathcal{C}}$, written as, $\langle f_{\phi, \theta}(x)| \hat{\mathcal{C}} | f_{\phi, \theta}(x) \rangle$. Out of many available observables, the simplest is the magnetization of a single qubit $j$, $\langle Z_j \rangle$, which represents the function in range $[-1, 1]$. This choice requires rescaling for other intervals. Other options are total magnetization in the system $\hat{\mathcal{C}} = \sum_j Z_j$. 

In addition to computing expectations, a weighted sum of observables is also possible, such as,
\begin{equation}
    \hat{\mathcal{C}} = \sum_{\ell} \alpha_{\ell} \hat{\mathcal{C}}_{\ell},
\end{equation}
where $\alpha_{\ell} \in \mathbb{R}$ are trainable coefficients, and $\hat{\mathcal{C}}_{\ell}$ are usual observables as described above. These tunable coefficients $\alpha_{\ell}$ are adjusted by optimizers such as gradient descent, Adam, etc. This is the beauty of hybrid quantum-classical algorithms.

\subsection{Loss Function Handling}

The loss function to solve differential equations using quantum circuits is similar to classical PINNs. The classical optimizer updates the variational parameters such that the predicted function matches the desired function. This is achieved by reducing the distance between differential expression and zero evaluated at a set of collocation points. It is also required to satisfy the initial and boundary conditions. In other words, this resembles an optimization problem for a loss function of derivatives and functions evaluated at collocation points.

A generalized loss function parameterized by variational parameters $\theta$ is defined as,
\begin{equation}\label{eq:5.23}
    \mathcal{L}_{\theta}[d_xf, f, x] = \mathcal{L}_{\theta}^{(\text{diff})}[d_xf, f, x] + \mathcal{L}_{\theta}^{(\text{boundary})}[f, x],
\end{equation}
where $\mathcal{L}_{\theta}^{(\text{diff})}$ corresponds to differentials, and $\mathcal{L}_{\theta}^{(\text{boundary})}$ corresponds to boundary conditions. The differential loss is defined as,
\begin{equation}\label{eq:5.24}
    \mathcal{L}_{\theta}^{(\text{diff})}[d_xf, f, x] = \frac{1}{M} \sum_{i=1}^M L(F[d_xf(x_i), f(x_i), x_i], 0),
\end{equation}
where $L(a, b)$ is a function to describe the distance between the two arguments $a$ and $b$. The loss is evaluated at $M$ points and normalized by the size $M$. Functional $F$ determines the differential equation described in the form $F[d_xf, f, x] = 0$. Here, the functional means it includes all differential equations when the problem is a system, such that all equations are taken into account. For example, a Neumann boundary condition can also be included in the system. The boundary conditions especially, Dirichlet boundary conditions loss are defined by,
\begin{equation}\label{eq:5.25}
    \mathcal{L}_{\theta}^{(\text{boundary})}[f, x] = \eta L(f(x_0), u_0),
\end{equation}
which calculates the distance between the function value at the boundary $x_0$ and the given boundary value $u_0$. Similarly, the point $x_0$ here also can be the initial point instead of the boundary value. The factor $\eta$ is referred to as the boundary pinning coefficient which regulates the weight of the boundary term in the optimization procedure. To emphasize the boundary precision, larger $\eta > 1$ can be used to prioritize.

The function $L$ is for distance definition and can be defined in many ways. Popular one is a \textit{mean squared error} (MSE) given as,
\begin{equation}
    L(a, b) = (a - b)^2.
\end{equation}

Additionally, the other distance choice is \textit{mean absolute error} (MAE), defined as, $L(a, b) = |a - b|$. The choice of loss function affects the optimizer's performance on convergence. MSE works better by punishing the loss function harder with larger distances and lighter with smaller distances. MAE converges slowly, but once the training is close to the optimal solution, it can achieve higher accuracy than MSE.

\subsection{Floating Boundary Handling}\label{sec:floating}
A simple way of handling Dirichlet boundary or initial conditions would be adding the boundary loss terms (\cref{eq:5.25}) into the total loss function (\cref{eq:5.23}) and trying to optimize both functions simultaneously. In multiple loss functions, trying to optimize one loss function will affect the performance of the other. However, if the differential equation has the term $f(x)$ in it, there is a possibility to solve the problem in a more efficient way known as \textbf{floating boundary handling}. 

Floating boundary handling corresponds to recursively shifting the predicted solution based on the boundary or initial point. With this method, the boundary loss term can be removed from the total loss, rather it is encoded in the expectation of observable. As the function is parameterized to match a specific boundary, boundary information is contained within the function and its derivatives. In this way, the need for separate boundary loss terms can be eradicated. The updated function is described as,
\begin{equation}\label{eq:5.27}
    f(x) = f_b + \langle f_{\phi, \theta}(x)|\hat{\mathcal{C}} |f_{\phi, \theta}(x) \rangle,
\end{equation}
where $f_b \in \mathbb{R}$ is the parameter that is adjusted after each iteration step as,
\begin{equation}\label{eq:5.28}
    f_b = u_0 - \langle f_{\phi, \theta}(x_0)|\hat{\mathcal{C}} |f_{\phi, \theta}(x_0) \rangle.
\end{equation}

The solver will effectively find a function $\langle f_{\phi, \theta}(x)|\hat{\mathcal{C}} |f_{\phi, \theta}(x) \rangle$ that solves the differential equation shifted to any position, then being shifted to the desired initial condition as shown in \cref{eq:5.28}. The advantage of this method is that the derivative loss term does not need to compete with the boundary loss. This also allows the optimizer to find the optimal angles and reduces the dependence on initial $\theta_{\text{init}}$. This method can be generalized to multivariable problems which have an initial condition as a function of a subset of the independent variables.

\end{document}